\documentclass[conference,10pt]{IEEEtran}
\IEEEoverridecommandlockouts
\usepackage{amsfonts}
\usepackage{amsmath}
\usepackage{amssymb} 
\usepackage{array}
\usepackage{bm} 
\usepackage{bbm}
\usepackage{color} 
\usepackage{cases} 
\usepackage{cite}
\usepackage{dsfont}
\usepackage[bookmarks=false]{hyperref}
\hypersetup{
	colorlinks=true,
	linkcolor=blue,
	filecolor=blue,
	citecolor = blue,      
	urlcolor=cyan,}
\usepackage[pdftex]{graphicx}
\DeclareGraphicsExtensions{.eps,.pdf,.png,.jpg,.gif,.jpeg,.pstex}
\usepackage{hhline}
\usepackage{multirow}
\usepackage{makecell}
\usepackage{stfloats}
\usepackage{textcomp}
\usepackage{url}
\usepackage{units} 
\usepackage{verbatim}

\usepackage[caption=false, font=footnotesize]{subfig}
\usepackage[printonlyused]{acronym}
\usepackage[table]{xcolor}

\usepackage{algorithm}
\usepackage{algpseudocode}
\usepackage{tikz} 
\usepackage[utf8]{inputenc}
\usepackage{pgfplots} 
\pgfplotsset{width=10cm,compat=1.9}
\usepackage{pgfplotstable}
\usepackage{xurl} 
\usepackage[colorinlistoftodos,bordercolor=orange,backgroundcolor=orange!20,linecolor=orange,textsize=scriptsize]{todonotes}
\usepackage{yfonts}

\algtext*{EndWhile}
\algtext*{EndIf}
\algtext*{EndFor}
\usetikzlibrary {arrows.meta}
\makeatletter

\makeatletter
\algnewcommand{\LineComment}[1]{\Statex \hskip\ALG@thistlm \(\triangleright\) #1}
\makeatother

\newacro{aoa} [AoA] {angle-of-arrival}
\newacro{aod} [AoD] {angle-of-departure}
\newacro{amp} [AMP] {approximate message passing}
\newacro{arv} [ARV] {array response vector}
\newacro{awgn} [AWGN] {additive white Gaussian noise}
\newacro{bs} [BS] {base station}
\newacro{cdf} [CDF] {cumulative distribution function}
\newacro{cf} [CF] {cell-free}
\newacro{ckm} [CKM] {channel knowledge map}
\newacro{cp} [CP] {cyclic prefix}
\newacro{crb} [CRB] {Cram{\'e}r-Rao bound}
\newacro{crlb} [CRLB] {Cram{\'e}r-Rao lower bound}
\newacro{cs} [CS] {compressed sensing}
\newacro{csi} [CSI] {channel state information}
\newacro{dft}[DFT]{discrete Fourier transform}
\newacro{em} [EM] {electromagnetic}
\newacro{eer} [EER] {equal error rate}
\newacro{fa} [FA] {false alarm}
\newacro{fim} [FIM] {Fisher information matrix}
\newacro{ghz} [GHz] {gigahertz}
\newacro{glrt} [GLRT]{generalized likelihood ratio test}
\newacro{iid}[i.i.d.]{independently and identically distributed}
\newacro{isac} [ISAC] {integrated sensing and communications}
\newacro{kld} [KLD] {Kullback–Leibler divergence}
\newacro{llr}[LLR]{log-likelihood ratio}
\newacro{lmmse} [LMMSE] {linear minimum mean-square error}
\newacro{los} [LoS] {line-of-sight}
\newacro{ls} [LS] {least-square}
\newacro{lse} [LSE] {least-square estimator}
\newacro{lsfc} [LSFC] {large scale fading coefficient}
\newacro{lb} [LB] {lower bound}
\newacro{map} [MAP] {maximum a posteriori}
\newacro{mc} [MC] {mutual coupling}
\newacro{md} [MD] {misdetection}
\newacro{mf} [MF] {matched filter}
\newacro{ml} [ML] {maximum likelihood}
\newacro{mle} [MLE] {maximum likelihood estimation}
\newacro{mmle} [MMLE] {mismatched maximum likelihood estimation}
\newacro{mcrb} [MCRB] {misspecified Cram\'er-Rao bound}
\newacro{mcrlb} [MCRLB] {misspecified Cram\'er-Rao lower bound}
\newacro{mse} [MSE] {mean-square error}
\newacro{mimo} [MIMO] {multiple-input multiple-output}
\newacro{mmwave} [mmWave] {millimeter-wave}
\newacro{mmse} [MMSE] {minimum mean-square error}
\newacro{nlos} [NLoS] {non-line-of-sight}
\newacro{nmse} [NMSE] {normalized mean squared error}
\newacro{np} [NP] {Neyman-Pearson}
\newacro{omp} [OMP] {orthogonal matching pursuit}
\newacro{ofdm} [OFDM] {orthogonal frequency-division multiplexing}
\newacro{pdf}[PDF]{probability density function}
\newacro{rv} [RV] {random variable}
\newacro{rach} [RACH] {random access channel}
\newacro{rm} [RM]{radio map}
\newacro{rss} [RSS] {received signal strength}
\newacro{ru} [RU] {radio unit}
\newacro{ris} [RIS] {Reconfigurable intelligent surface}
\newacro{simo} [SIMO] {single-input multiple-output}
\newacro{se} [SE] {state evolution}
\newacro{svd} [SVD] {singular value decomposition}
\newacro{snr} [SNR] {signal-to-noise ratio}
\newacro{tdd} [TDD] {time division duplexing}
\newacro{tdoa} [TDoA] {time-difference-of-arrival}
\newacro{toa} [ToA] {time-of-arrival}
\newacro{thz} [THz] {terahertz}
\newacro{ue} [UE] {user equipment}
\newacro{ula} [ULA] {uniform linear array}
\newacro{upa} [UPA] {uniform planar array}
\newacro{ura} [uRA] {unsourced random access}
\newacro{zc} [ZC] {Zadoff--Chu}

\newcommand{\Tran}{{\sf T}}
\newcommand{\Herm}{{\sf H}}

\newcommand{\abs}[1]{\left|{#1}\right|}

\newcommand{\diag}{{\hbox{diag}}}

\newcommand{\nth}[1]{{#1}{\text{th}}}

\renewcommand{\vec}{{\rm vec}}

\newcommand{\az}{\mathrm{az}}

\newcommand{\el}{\mathrm{el}}

\newcommand{\MC}{\mathrm{MC}}
\newcommand{\CV}{\mathrm{CV}}
\newcommand{\NEUM}{\mathrm{neu}}
\newcommand{\APR}{\mathrm{apr}}
\newcommand{\PROP}{\mathrm{prop}}
\newcommand{\SCN}{\mathrm{SCN}}
\newcommand{\TMP}{\mathrm{TMP}}
\newcommand{\THR}{\mathrm{THR}}

\newcommand{\av}{{\bf a}}

\newcommand{\cv}{{\bf c}}

\newcommand{\fv}{{\bf f}}

\newcommand{\nv}{{\bf n}}

\newcommand{\pv}{{\bf p}}

\newcommand{\rv}{{\bf r}}

\newcommand{\wv}{{\bf w}}

\newcommand{\xv}{{\bf x}}
\newcommand{\yv}{{\bf y}}

\newcommand{\Am}{{\bf A}}
\newcommand{\Bm}{{\bf B}}

\newcommand{\Dm}{{\bf D}}

\newcommand{\Gm}{{\bf G}}
\newcommand{\Hm}{{\bf H}}
\newcommand{\Id}{{\bf I}}

\newcommand{\Pm}{{\bf P}}

\newcommand{\Sm}{{\bf S}}

\newcommand{\Ym}{{\bf Y}}

\newcommand{\Irm}{{\rm I}}

\newcommand{\Rrm}{{\rm R}}

\newcommand{\Trm}{{\rm T}}

\newcommand{\gammav}{\hbox{\boldmath$\gamma$}}

\newcommand{\phiv}{\hbox{\boldmath$\phi$}}
\newcommand{\varphiv}{\hbox{\boldmath$\varphi$}}
\newcommand{\psiv}{\hbox{\boldmath$\psi$}}
\newcommand{\thetav}{\hbox{$\boldsymbol\theta$}}
\newcommand{\varthetav}{\hbox{$\boldsymbol\vartheta$}}

\newcommand{\omegav}{\hbox{\boldmath$\omega$}}
\newcommand{\xiv}{\hbox{\boldmath$\xi$}}
\newcommand{\sigmav}{\hbox{\boldmath$\sigma$}}

\newcommand{\Gammam}{\hbox{\boldmath$\Gamma$}}

\newcommand{\Sigmam}{\hbox{\boldmath$\Sigma$}}

\newcommand{\Psim}{\hbox{\boldmath$\Psi$}}

\newcommand{\Thetam}{\hbox{\boldmath$\Theta$}}
\newcommand{\Omegam}{\hbox{\boldmath$\Omega$}}
\newcommand{\Xim}{\hbox{\boldmath$\Xi$}}

\newfont{\bb}{msbm10 scaled 1100}
\newcommand{\CC}{\mbox{\bb C}}

\newcommand{\RR}{\mbox{\bb R}}

\newcommand{\EE}{\mbox{\bb E}}

\makeatother

\makeatletter
\def\ps@IEEEtitlepagestyle{%
  \def\@oddfoot{\mycopyrightnotice}%
  \def\@oddhead{\hbox{}\@IEEEheaderstyle\leftmark\hfil\thepage}\relax
  \def\@evenhead{\@IEEEheaderstyle\thepage\hfil\leftmark\hbox{}}\relax
  \def\@evenfoot{}%
}

\def\mycopyrightnotice{%
  \begin{minipage}{\textwidth}
  \centering \scriptsize
    Copyright~\copyright~2026 IEEE. Personal use of this material is permitted. Permission from IEEE must be obtained for all other uses, in any current or future media, including reprinting/republishing this material for advertising or promotional purposes, creating new collective works, for resale or redistribution to servers or lists, or reuse of any copyrighted component of this work in other works.
    
    Accepted for publication in the IEEE NextGCom 2026.
  \end{minipage}
}
\makeatother

\begin{document}

\bstctlcite{IEEEexample:BSTcontrol}

\title{Exploiting Mutual Coupling Structure for Channel Estimation of Active RIS-Assisted Links\\
\thanks{The work of S. Tarboush was supported by the European Union, through the Horizon Europe Marie Skłodowska-Curie Doctoral Networks Programme “Intelligent sensing and communication as training network for perceptive mobile networks in 6G (ISAC-NEWTON)” under Grant 101169496. The work of G. Caire was supported by the BMFTR Germany in the program of “Souverän. Digital. Vernetzt.” Joint Project 6G-RIC (Project IDs 16KISK030). The work of T.~Y.~Al-Naffouri was supported by the KAUST Office of Sponsored Research (OSR) under Award No. ORFS-CRG12-2024-6478. The work of H. Sarieddeen was supported by the AUB University Research Board (URB) and Vertically Integrated Projects (VIP) program.
}
}

\author{
    \IEEEauthorblockN{
        Simon Tarboush\IEEEauthorrefmark{1},
        Pinjun Zheng\IEEEauthorrefmark{2},
        Hadi~Sarieddeen\IEEEauthorrefmark{3},
        Tareq~Y.~Al-Naffouri\IEEEauthorrefmark{4},
		and Giuseppe Caire\IEEEauthorrefmark{1}
        }
	\IEEEauthorblockA{
        \IEEEauthorrefmark{1}\small Communications and Information Theory Chair, Faculty of Electrical Engineering and Computer Science,\\ Technische Universit{\"a}t Berlin, 10587 Berlin, Germany.
        }
    \IEEEauthorblockA{
        \IEEEauthorrefmark{2}\small School of Engineering, University of British Columbia, Kelowna, BC V1V 1V7, Canada.
        }
    \IEEEauthorblockA{
        \IEEEauthorrefmark{3}\small Electrical and Computer Engineering Department, American University of Beirut (AUB), Lebanon.
        }
    \IEEEauthorblockA{
        \IEEEauthorrefmark{4}\small Electrical and Computer Engineering Program, Division of Computer, Electrical and Mathematical Sciences and Engineering (CEMSE),\\ King Abdullah University of Science and Technology (KAUST), Thuwal, 23955-6900, Kingdom of Saudi Arabia.
        \\
        Emails: (simon.tarboush; caire@tu-berlin.de), pinjun.zheng@ubc.ca, hadi.sarieddeen@aub.edu.lb, tareq.alnaffouri@kaust.edu.sa.
        }
}

\maketitle

\begin{abstract}
Accurate channel modeling and estimation of active reconfigurable intelligent surface (RIS)-assisted links with densely integrated elements are essential to fully unleashing this technology's potential. This work adopts a physically consistent model incorporating mutual coupling (MC) effects, modeled via scattering parameters, in RIS-aided communication. We formulate the MC-aware channel estimation as a compressed sensing (CS) problem. The MC effect leads to an increase in the sensing matrix dimensions. This increased dimensionality substantially elevates the complexity of the formulated CS problem. To overcome this, we propose a low-complexity estimator that leverages the structure of the scattering matrix and MC mechanisms to obtain a reduced-size design sensing matrix. Numerical results demonstrate that our approach outperforms MC-unaware estimators by several dBs, achieving accuracy comparable to fully MC-aware solutions but with significantly lower complexity.
\end{abstract}

\begin{IEEEkeywords}
Reconfigurable intelligent surface, physically consistent model, mutual coupling, scattering parameters.
\end{IEEEkeywords}

\section{Introduction}
\label{sec:intro}

\ac{ris} offers new functionalities to dynamically shape radio propagation channels and manipulate \ac{em} waves to enhance future wireless networks~\cite{Di2020Smart,Wang2026Millimeter}, particularly in promising spectra like the upper mid-band, millimeter-wave frequencies, and the terahertz band~\cite{Kang2024Cellular,Tarboush2021Teramimo,Sarieddeen2021Overview}. \ac{ris} consists of a planar metamaterial-based surface with a large number of nearly passive reflecting elements. At high frequencies, the severe fading and limited capacity gains are the major drawbacks of passive \ac{ris}. Thus, active~\ac{ris} has been proposed to mitigate these limitations and can be realized using reflection-type amplifiers integrated into the~\ac{ris} unit cells~\cite{Zhang2022Active}.

Most \ac{ris} models assume a diagonal reflection matrix~\cite{Pan2022An,Swindlehurst2022Channel}, neglecting the \ac{mc} effect caused by \ac{em} interactions among closely spaced cells. However, \ac{mc} is inevitable in densely packed surfaces~\cite{Gradoni2021End,Akrout2023Physically} and is further exacerbated by the high amplification of active \ac{ris}~\cite{Zheng2024On}. Under these conditions, the \ac{mc} significantly affects both the \ac{ris} unit cell impedance and radiation pattern. Accurate modeling of \ac{mc} should integrate physically consistent \ac{em} propagation laws. To this end, several works introduce frameworks that account for \ac{mc} in RIS-aided links~\cite{Gradoni2021End,Akrout2023Physically,Zheng2024On}. Among these frameworks, models based on the scattering matrix (S-parameters)~\cite{Shen2022Modeling,Nerini2024Universal,Renzo2024Electromagnetic,Dkhan2025RIS} have the advantage of being verifiable through practical measurements~\cite{Zheng2024Mutual}. Conventional \ac{mc}-unaware \ac{ris} channel estimation has been discussed extensively, modeling the estimation problem as a sparse recovery problem that can be solved through standard~\ac{cs} techniques~\cite{Wang2020Compressed,Swindlehurst2022Channel,Iian2022Reconfigurable,Zheng2022Survey}. However, accurate \ac{ris} channel estimation and beamforming inherently demand physically consistent, \ac{mc}-aware models~\cite{Zheng2024On,Wijekoon2025Physically,Abrardo2024Design,Zheng2026Mutual}.

Our previous work~\cite{Zheng2026Mutual} demonstrates that the \ac{ris}-assisted channel estimation problem with \ac{mc} can still be formulated and solved using \ac{cs} techniques. However, incorporating \ac{mc} introduces a dimension lift problem, significantly increasing the scale of the formulated \ac{cs} problem. According to~\cite{Zheng2026Mutual}, a dimensionality comparison between the conventional \ac{mc}-unaware and the exact \ac{mc}-aware \ac{cs} formulations is
\begin{align}
    \label{eq:intro_y_nomc}\text{MC-unaware CS: } \yv&=\underbrace{\Psim_\CV}_{M\times K N_\Irm}\underbrace{\Dm_\CV}_{\ K N_\Irm\times G }\xv,\\
    \label{eq:intro_y_mc}\text{MC-aware CS: } \yv&=\underbrace{\Psim_\MC}_{M\times KN_\Irm^2}\underbrace{\Dm_\MC}_{\ KN_\Irm^2\times G^2}\xv,
\end{align}
where~$\Psim$ and $\Dm$ denote the measurement and dictionary matrices, respectively, and~$N_\mathrm{I}$ is the number of \ac{ris} unit cells. Specifically, by incorporating \ac{ris} \ac{mc}, a scale increase from~$N_\Irm$ to~$N_\Irm^2$ occurs in one dimension for both the measurement and dictionary matrices, and the dictionary matrix columns also expand from~$G$ to~$G^2$. This occurs because, when considering \ac{mc}, we must not only include the reflection coefficients of~$N_\Irm$ unit cells but also account for the coupling effects between each pair of these unit cells. Consequently, $\mathcal{O}(N_\Irm^2)$ parameters are required to characterize this process. This dimension lift can significantly increase the time complexity of the formulated \ac{cs} problem, particularly in large-scale \ac{ris} scenarios.
To address this problem, \cite{Zheng2026Mutual} proposes a two-step method that reduces the dictionary size~$G^2$ in~$\Dm_\MC$. However, this method relies on the \ac{mc}-unaware model to get a coarse estimate that fails under strong \ac{mc} scenarios.

In this work, we propose novel reduced-size yet accurate measurement and dictionary matrices inspired by the \ac{mc} structure. By leveraging the structure of the scattering matrix and using Neumann series expansion, we construct a \ac{mc}-aware dictionary with $\mathcal{O}(N_\Irm)$ rows instead of $N_\Irm^2$, correspondingly reducing the measurement matrix columns. This is different from our previous proposal in~\cite{Zheng2026Mutual} which only reduces the dictionary matrix columns. 
Compared to \cite{Zheng2026Mutual}, the proposed solution offers two distinct contributions: (i) it achieves dimensionality reduction by exploiting the inherent sparsity within the physical mechanism of \ac{mc}, without relying on prior knowledge of the support set as required in~\cite{Zheng2026Mutual}; and (ii) instead of reducing the number of atoms in the dictionary, as done in~\cite{Zheng2026Mutual}, this method minimizes the length of each atom. Numerical evaluations confirm that this approach surpasses conventional \ac{mc}-unaware models and achieves accuracy comparable to the high-complexity exact model while reducing complexity, even under strong \ac{mc}.

{\bf Notation:} Non-bold lower and upper case letters (e.g., $a, A$) denote scalars, bold lower case letters (e.g., $\av$) denote vectors, and bold upper case letters (e.g., $\Am$) denote matrices.  For two $M\times N$ matrices $\Am$ and $\Bm$, $\Am\otimes\Bm$ denotes the $M^2\times N^2$ Kronecker product matrix, $\Am\bullet\Bm$ is the $M\times N^2$ row-wise Khatri-Rao product matrix, and $\odot$ is the Hadamard product.

\section{System and Channel Modeling}
\label{sec:Ch_Model}
Consider a narrow-band far-field high-frequency \ac{tdd} uplink \ac{mimo} communication system, consisting of an~$N_\Trm$-antenna \ac{ue}, an~$N_\Rrm$-antenna \ac{bs}, and an~$N_\Irm$-unit cell active \ac{ris}, where the \ac{ue}-\ac{bs} direct link is assumed blocked. The \ac{ue} deploys a \ac{upa} consisting of $N_\Trm\!=\!N_\Trm^h\!\times\! N_\Trm^v$ antenna elements, where the superscripts $h$ and $v$ denote the horizontal and vertical dimensions, respectively. The same assumptions are applied for the \ac{bs} and \ac{ris}, i.e., $N_\Rrm=N_\Rrm^h\times N_\Rrm^v$ and $N_\Irm=N_\Irm^h\times N_\Irm^v$, respectively. By utilizing S-parameter-based multiport network theory, the \ac{mc}-aware channel model is~\cite{Shen2022Modeling,Zheng2024Mutual,Nerini2024Universal,Renzo2024Electromagnetic,Zheng2026Mutual}\footnote{An experimental validation  of~\eqref{eq:MC_ChM} using real \ac{ris} prototype has been conducted in~\cite{Zheng2024Mutual}. Models using multiport network theory with impedance matrix (Z-parameters) or admittance matrix (Y-parameters)~\cite{Gradoni2021End,Akrout2023Physically} are equivalent to those using S-parameters~\cite{Nerini2024Universal,Renzo2024Electromagnetic}.}
\begin{equation}
    \label{eq:MC_ChM}
    \Hm_\MC = \Hm_{\Rrm\Irm} (\Gammam^{-1} - \Sm)^{-1} \Hm_{\Irm\Trm},
\end{equation}
where~$\Hm_{\Rrm\Irm}\in\CC^{N_\Rrm \times N_\Irm}$ and~$\Hm_{\mathrm{IT}}\in\CC^{N_\Irm\times N_\Trm}$ are the individual \ac{ris}-\ac{bs} and \ac{ue}-\ac{ris} subchannels,~$\Gammam=\diag(\gammav)$ is the \ac{ris} reflection matrix following~$\gammav\in\CC^{N_\Irm}$. The reflection coefficient of the~$\nth{i}$ \ac{ris} unit cell is $\gamma_i=a_ie^{j\nu_i}$, where~$a_i>1$ represents the active \ac{ris} element amplification factor and~$\nu_i\in[0,2\pi)$ denotes its phase shift. The scattering matrix~$\Sm\in\CC^{N_\Irm\times N_\Irm}$ models the \ac{mc} effect between the \ac{ris} unit cells. The $\nth{(r,c)}$ entry, $[\Sm]_{r,c}, (1\leq \{r,c\} \leq N_\Irm)$, represents the scattering parameter between the~$\nth{r}$ and~$\nth{c}$ unit cells of the \ac{ris}, and is defined by \cite{Shen2022Modeling}
\begin{equation}
	[\Sm]_{r,c} = \frac{V_r^\mathrm{out}}{V_c^\mathrm{in}}\bigg|_{V_k^\mathrm{in}=0,\ \forall k\neq c},
\end{equation}
where~$V_c^\mathrm{in}/V_r^\mathrm{out}$ denotes the voltage wave incident to/reflected from the~$\nth{c}/\nth{r}$ unit cell. The conventional \ac{ris}-assisted channel model assumes that each \ac{ris} unit cell reflects incident \ac{em} waves independently without any interactions, and can be obtained from~\eqref{eq:MC_ChM} by setting~$\Sm = \mathbf{0}$~\cite{Pan2022An,Swindlehurst2022Channel,Wang2020Compressed,Iian2022Reconfigurable,Zheng2022Survey} so that
\begin{equation}
\label{eq:convCHM}
	\Hm_\CV = \Hm_{\Rrm\Irm} \Gammam \Hm_{\Irm\Trm}.
\end{equation}
For simplicity, we refer to~\eqref{eq:MC_ChM} as the \emph{exact model} and~\eqref{eq:convCHM} as the \emph{conventional model} throughout the paper.

We assume that the \ac{ue}, \ac{bs}, and \ac{ris} deploy a \ac{upa} geometry on the YZ-plane of their body coordinate system; then the \ac{arv} can be expressed as~\cite{Tarboush2023Compressive}
\begin{equation}
    \label{eq:UPA_ARV}
    \av_\varsigma(\psiv_\ell)=\frac{1}{\sqrt{N_\varsigma}}e^{-j2\pi\psi_\ell^h\nv(N_\varsigma^h)}\otimes e^{-j2\pi\psi_\ell^v\nv(N_\varsigma^v)},
\end{equation}
where $\varsigma\in\{\Trm,\Rrm,\Irm\}$,~$\nv(N)=[0,1,\dots,N-1]^\Tran$, $\psiv_\ell=[\psi_\ell^\az,\psi_\ell^\el]^\Tran$, for the $\nth{\ell}$ channel path, comprises azimuth and elevation components and represents \ac{aod} or \ac{aoa}. The spatial angles are $\psi_\ell^h\triangleq{d_\varsigma}\sin(\psi_\ell^\az)\sin(\psi_\ell^\el)/{\lambda_c}$ and~$\psi_\ell^v\triangleq{d_\varsigma}\cos(\psi_\ell^\el)/{\lambda_c}$, respectively, with $d_\varsigma$ being the inter-element spacing of the array, and~$\lambda_c$ being the wavelength for an operating frequency~$f_c$.

The frequency-domain \ac{ue}-\ac{ris} channel response and its beamspace representation, following~\eqref{eq:UPA_ARV}, are~\cite{Swindlehurst2022Channel,Pan2022An,Wang2020Compressed}
\begin{align}
\label{eq:ch_TxRIS_spatial_domain}
 \Hm_{\Irm\Trm} &= \sqrt{\frac{N_\Irm N_\Trm}{L_\Trm}}\sum_{\ell=1}^{L_{\Trm}} \alpha_\ell\av_\Irm(\phiv_\ell)\av^\Tran_\Trm(\varphiv_\ell),\\
 &\label{eq:ch_TxRIS_angle_domain}= \Am_\Irm(\phiv)\Sigmam_{\Irm\Trm}\Am^\Tran_\Trm(\varphiv),
\end{align} 
where~$L_{\Trm}$ denotes the number of paths between the \ac{ue} and \ac{ris}, where~$\ell\!=\!1$ stands for the \ac{los} path and~$\ell=2,\dots,L_\Trm$ correspond to $L_\Trm\!-\!1$ \ac{nlos} paths,~$\alpha_\ell\in\CC$ denotes the complex channel gain,~$\varphiv_\ell\in\RR^2$ the \ac{aod} at the \ac{ue},~$\phiv_\ell\in\RR^2$ the \ac{aoa} at the \ac{ris}, and~$\av_\Trm(\varphiv_\ell)\in\CC^{N_\Trm}$ and~$\av_\Irm(\phiv_\ell)\in\CC^{N_\Irm}$ are the \ac{arv} at the \ac{ue} and \ac{ris}, following~\eqref{eq:UPA_ARV}, corresponding to~$\varphiv_\ell$ and~$\phiv_\ell$, respectively. In~\eqref{eq:ch_TxRIS_angle_domain}, $\Sigmam_{\Irm\Trm}=\sqrt{N_\Irm N_\Trm/{L_\Trm}} \diag(\alpha_1,\alpha_2,\cdots,\alpha_{L_\Trm})\in\CC^{L_\Trm \times L_\Trm}$ is the beamspace channel matrix, and~$\Am_\Trm(\varphiv)=[\av_{\Trm}(\varphiv_1),\av_{\Trm}(\varphiv_2),\cdots,\av_\Trm(\varphiv_{L_\Trm})]\in\CC^{N_\Trm\times L_\Trm}$,~$\Am_\Irm(\phiv)=[\av_{\Irm}(\phiv_1),\av_{\Irm}(\phiv_2),\cdots,\av_\Irm(\phiv_{L_\Trm})]\in\CC^{N_\Irm\times L_\Trm}$ are the array response matrices, after concatenating all the~\acp{arv} in~\eqref{eq:ch_TxRIS_spatial_domain} with the angle vectors defined $\varphiv=[\varphiv_1^\Tran,\varphiv_2^\Tran,\dots,\varphiv_{L_\Trm}^\Tran]^\Tran\in\RR^{2L_\Trm}$ and $\phiv=[\phiv_1^\Tran,\phiv_2^\Tran,\dots,\phiv_{L_\Trm}^\Tran]^\Tran\in\RR^{2L_\Trm}$. Following the same steps and definitions of~\eqref{eq:ch_TxRIS_spatial_domain} and \eqref{eq:ch_TxRIS_angle_domain}, and by applying proper notations, the \ac{ris}-\ac{bs} channel response and beamspace are
\begin{align}
    \label{eq:ch_RISRx_spatial_domain}
    \Hm_{\Rrm\Irm} &= \sqrt{\frac{N_\Irm N_\Rrm}{L_{\Rrm}}}\sum_{\ell=1}^{L_{\Rrm}} \rho_\ell\av_\Rrm(\varthetav_\ell)\av_\Irm^\Tran(\thetav_\ell),\\
    &\label{eq:ch_RISRx_angle_domain}    =\Am_\Rrm(\varthetav)\Sigmam_{\Rrm\Irm}\Am_\Irm^\Tran(\thetav).
\end{align}

\section{MC-Aware Sparse Recovery Formulation}
\label{sec:sparse_rec}

\begin{table*}[htb!]
    \centering
    \small
    \caption{Complexity analysis of required measurement and dictionary matrices for \ac{mc}-unaware and \ac{mc}-aware \ac{cs} models}
    \begin{tabular}{|c|c|c|c|c|c|c|c|c|}
    \hline
    & \multicolumn{4}{|c|}{Theory} & \multicolumn{4}{|c|}{\makecell{$N_\Trm\!=G_\Trm\!=\!2\times2, N_\Rrm\!=G_\Rrm\!=\!4\times4,$ \\ $N_\Irm\!=G_\Irm\!=\!16\times16, M_\Rrm\!=\!N_\Rrm/2, M_\Irm\!=\!N_\Irm/2$}} \\ \cline{2-9}
    &\multicolumn{2}{|c|}{$\Psim_\chi$} & \multicolumn{2}{|c|}{$\bar\Dm_\chi$} & \multicolumn{2}{|c|}{$\Psim_\chi$} & \multicolumn{2}{|c|}{$\bar\Dm_\chi$} \\ \hline
    & \# rows & \# columns & \# rows & \# columns & \# rows & \# columns & \# rows & \# columns \\ \hline
    \ac{mc}-unaware & $M_\Rrm M_\Irm$ & $N_\Rrm N_\Trm N_\Irm$ & $N_\Rrm N_\Trm N_\Irm$ & $G_\Rrm G_\Trm G_\Irm$ & 1024 & 16384 & 16384 & 16384 \\ \hline
    \ac{mc}-aware & $M_\Rrm M_\Irm$ & $N_\Rrm N_\Trm N_\Irm^2$ & $N_\Rrm N_\Trm N^2_\Irm$ & $G_\Rrm G_\Trm G^2_\Irm$ & 1024 & 4194304 & 4194304 & 4194304 \\ \hline
    \end{tabular}
    \label{table:complexity}
\end{table*}

The received signal $\yv\in\CC^{M_\Rrm M_\Irm}$ is obtained following a beam training procedure that involves sending pilots over training beams and adjusting the amplitudes and phase shifts of the \ac{ris} elements. The \ac{ris} utilizes~$M_\Irm$ random configurations $\gammav_{m_\Irm}\in\CC^{N_\Irm}$ \big(equivalently $\Gammam_{m_\Irm}=\diag(\gammav_{m_\Irm})$\big),~$m_\Irm=1,\dots,M_\Irm$, the \ac{ue} transmits~$M_\Rrm$ random precoding beams $\fv_{m_\Rrm}\in\CC^{N_\Trm}$, and the \ac{bs} records~$M_\Rrm$ received pilots through random combining beams $\wv_{m_\Rrm}\in\CC^{N_\Rrm}, \ m_\Rrm=1,\dots,M_\Rrm$. After collecting the measurements, the sparse formulation for the conventional/exact model is derived in \cite{Zheng2026Mutual} as
\begin{equation}
    \label{eq:Rxsig_CS}  \yv=\vec(\Ym)=\Xim_\chi\bar{\sigmav}_\chi+\vec(\Omegam),
\end{equation}
where $\Xim_\chi\!=\!\Psim_\chi\bar{\Dm}_\chi\!\in\!\mathbb{C}^{M_\Rrm M_\Irm \times \bar{G}_\chi}$ is the sensing matrix, $\bar{\sigmav}_\chi\!\in\!\CC^{\bar{G}_\chi}$ is a sparse vector, the cascaded channel is $\vec(\bar{\Gm}_\chi)=\Dm_\chi \bar{\sigmav}_\chi$, $\chi\in\{\CV,\MC\}$, and~$\Omegam\triangleq[\bar{\omegav}_1,\bar{\omegav}_2,\dots,\bar{\omegav}_{M_\Irm}]\in\CC^{M_\Rrm\times M_\Irm}$, $\bar{\omegav}_{m_\Irm}=[\omega_{1,m_\Irm},\omega_{2,m_\Irm},\dots,\omega_{M_\Rrm,m_\Irm}]^\Tran\in\CC^{M_\Rrm}$ with
\begin{equation}
    \omega_{m_\Rrm,m_\Irm} = \wv_{m_\Rrm}^\Herm\big(\Hm_{\Rrm\Irm}(\Gammam_{m_\Irm}^{-1} - \Sm)^{-1}\omegav^\Irm_{m_\Rrm,m_\Irm} + \omegav^\Rrm_{m_\Rrm,m_\Irm}\big),
\end{equation}
where~$\omegav^\Irm_{m_\Rrm,m_\Irm}\!\!\!\sim\!\mathcal{CN}(\mathbf{0},\!\sigma_\Irm^2\Id_{N_{\Irm}})$ and $\omegav^\Rrm_{m_\Rrm,m_\Irm}\!\!\!\!\sim\!\mathcal{CN}(\mathbf{0},\!\sigma_\Rrm^2\Id_{N_{\Rrm}})$ are the thermal noise at the active \ac{ris} and \ac{bs}, respectively. Moreover, the channels are approximated based on pre-determined dictionaries since we apply in this work an on-grid \ac{cs} technique to solve the problem. Following angular quantization of \eqref{eq:UPA_ARV} to a grid of size $G_\varsigma=G_\varsigma^h G_\varsigma^v$, where $\varsigma\in \{\Trm,\Rrm,\Irm\}$, the discrete array response matrices in \eqref{eq:ch_TxRIS_angle_domain} and \eqref{eq:ch_RISRx_angle_domain} are derived using $\tilde{\Am}_\varsigma(\psiv)\in\CC^{N_\varsigma\times G_i}$ with $\varsigma\in\{\Trm,\Rrm,\Irm\}$
\begin{equation}
    [\tilde{\Am}_\varsigma(\psiv)]_{:,g_{\varsigma}^h+(g_{\varsigma}^v\!-\!1)G_{\varsigma}^h} \!\!=\!\frac{1}{\sqrt{N_\varsigma}}e^{-j2\pi{\beta}^h_{g_{\varsigma}^h}\nv(\!N_\varsigma^h\!)}\!\!\!\otimes\!e^{-j2\pi{\beta}^v_{g_{\varsigma}^v}\nv(\!N_\varsigma^v\!)}\!\!,
\end{equation}
where the discrete spatial angles, $\varepsilon\in \{h,v\}$, are defined~\cite{Tarboush2023Compressive}
\begin{equation}      
    \label{eq:grid_spatialangle}
    {\beta}^\varepsilon_{g_\varsigma^\varepsilon}\!=\!\frac{2d_\varsigma}{\lambda_c G_\varsigma^\varepsilon}\!\Big(g_\varsigma^\varepsilon\!-\!\frac{G_\varsigma^\varepsilon\!+\!1}{2}\Big),\ g_\varsigma^\varepsilon\!=\!1,\!\cdots\!,G_\varsigma^\varepsilon;G_\varsigma^\varepsilon\!\geq\! N_\varsigma^\varepsilon.
\end{equation}

For the conventional model case in \eqref{eq:Rxsig_CS}, the measurement matrix is $\Psim_\CV\triangleq\Thetam_\CV^\Tran\otimes\Pm\in\CC^{M_\Rrm M_\Irm\times N_\Rrm  N_\Trm N_\Irm}$ with $\Pm=[\pv_1,\pv_2,\dots,\pv_{M_\Rrm}]^\Tran\in\CC^{M_\Rrm\times N_\Rrm N_\Trm}$, $\pv_{m_\Rrm}\triangleq\big(\fv^\Tran_{m_\Rrm}\otimes\wv_{m_\Rrm}^\Herm\big)^\Tran\in\CC^{N_\Rrm N_\Trm}$, and $\Thetam_\CV\triangleq{\left[{\gammav_1,\gammav_2,\cdots,\gammav_{M_\Irm}}\right]}\in\CC^{N_\Irm\times M_\Irm}$. The conventional dictionary matrix $\tilde{\Dm}_\CV \in \CC^{N_\Rrm N_\Trm N_\Irm \times G_\Rrm G_\Trm G^2_\Irm}$ is  
\begin{equation}
    \tilde{\Dm}_\CV = \tilde{\Am}_{\Irm\Irm}^{\CV}(\phiv,\thetav)\otimes \tilde{\Am}_{\Trm\Rrm}(\varphiv,\varthetav),\label{eq:Dtilde_cv}
\end{equation}
where $\tilde{\Am}_{\Irm\Irm}^{\CV}(\phiv,\thetav)\triangleq\big(\tilde{\Am}_\Irm(\phiv)\bullet\tilde{\Am}_\Irm(\thetav)\big)$ and $\tilde{\Am}_{\Trm\Rrm}(\varphiv,\varthetav) \triangleq\tilde{\Am}_\Trm(\varphiv)\otimes\tilde{\Am}_\Rrm(\varthetav)$. Following~\cite[Proposition~1]{Wang2020Compressed}, $\tilde{\Am}_{\Irm\Irm}^{\CV}(\phiv,\thetav)$ in \eqref{eq:Dtilde_cv} contains significant redundancy, with $G_\Rrm G_\Trm G_\Irm\!\leq\! \bar{G}_\CV\!\ll\! G_\Rrm G_\Trm G^2_\Irm$ denotes the number of distinct columns.

In the \ac{mc}-aware exact model, $\Psim_\MC\triangleq(\Thetam_\MC^\Tran\otimes\Pm)\in\CC^{M_\Rrm M_\Irm\times N_\Rrm N_\Trm N^2_\Irm}$ with $\Thetam_\MC\triangleq[\xiv_1, \xiv_2,\cdots,\xiv_{M_\Irm}]\in\CC^{N^2_\Irm\times M_\Irm}$, and $\xiv_{m_\Irm}\triangleq\vec(\bar{\Gammam}_{m_\Irm})\in\CC^{N^2_\Irm}$ with ~$\bar{\Gammam}_{m_\Irm}\triangleq(\Gammam_{m_\Irm}^{-1} - \Sm)^{-1}$ where due to the existence of matrix~$\Sm$, the \ac{ris} response~$\bar{\Gammam}_{m_\Irm}$ is no longer a diagonal matrix. Furthermore, the exact dictionary matrix $\tilde{\Dm}_\MC \in\CC^{N_\Rrm N_\Trm N^2_\Irm \times G_\Rrm G_\Trm G^2_\Irm}$ is
\begin{equation}
    \tilde{\Dm}_\MC = \tilde{\Am}_{\Irm\Irm}^\Tran(\phiv,\thetav) \otimes \tilde{\Am}_{\Trm\Rrm}(\varphiv,\varthetav),\label{eq:Dtilde_mc}
\end{equation}
where the Kronecker-based dictionary $\tilde{\Am}_{\Irm\Irm}(\phiv,\thetav)\triangleq\big(\tilde{\Am}_\Irm^\Tran(\phiv)\otimes\tilde{\Am}_\Irm^\Tran(\thetav)\big)$ does not have redundancy similar to $\bar{\Dm}_\CV$. We define~$\bar{\Dm}_\CV\in\CC^{N_\Rrm N_\Trm N_\Irm\!\times\! \bar{G}_\CV}$ and~$\bar{\Dm}_\MC\in\CC^{N_\Rrm N_\Trm N_\Irm^2\!\times\! \bar{G}_\MC}$, $\bar{G}_\MC=G_\Rrm G_\Trm G^2_\Irm$ as unique dictionaries that contain only distinct columns and are used in \eqref{eq:Rxsig_CS}. Table~\ref{table:complexity} compares the measurement and dictionary matrix sizes for conventional and exact models, with an illustrative numerical example. 

\section{Proposed MC-Aware Channel Estimation}
\begin{figure*}[htb]
 \centering
 \subfloat[Conceptual illustration of \ac{mc} effect when one unit cell is excited]{\label{fig:mc_mech} \includegraphics[width=0.46\linewidth,height=0.23\textwidth]{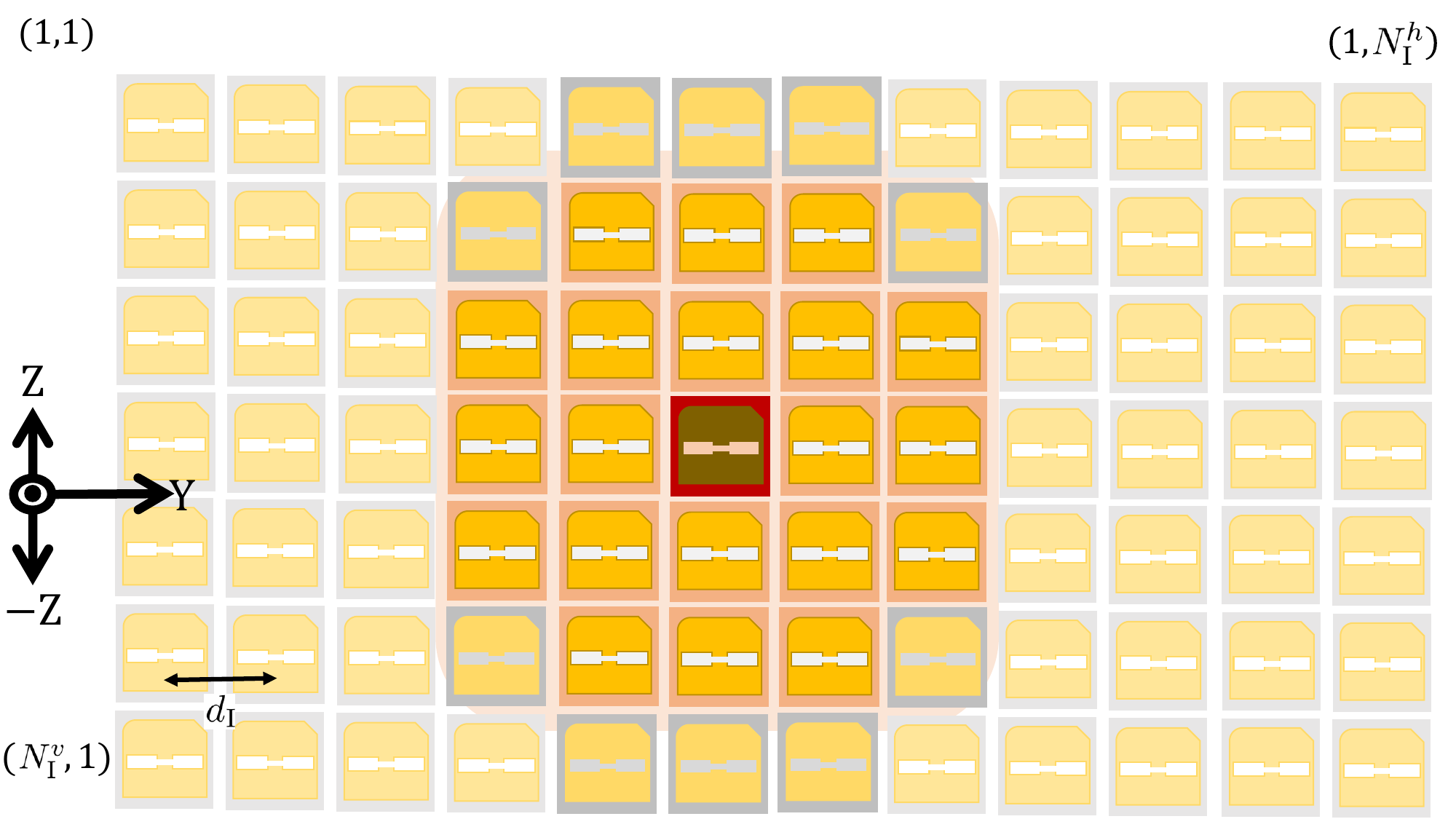}}
 \hfill
 \subfloat[Measured S-parameters, following~\cite{Zheng2024Mutual}, of real \ac{ris} prototype~\cite{Wang2024Wideband}]{\label{fig:real_s} \includegraphics[width=0.4\linewidth,height=0.23\textwidth]{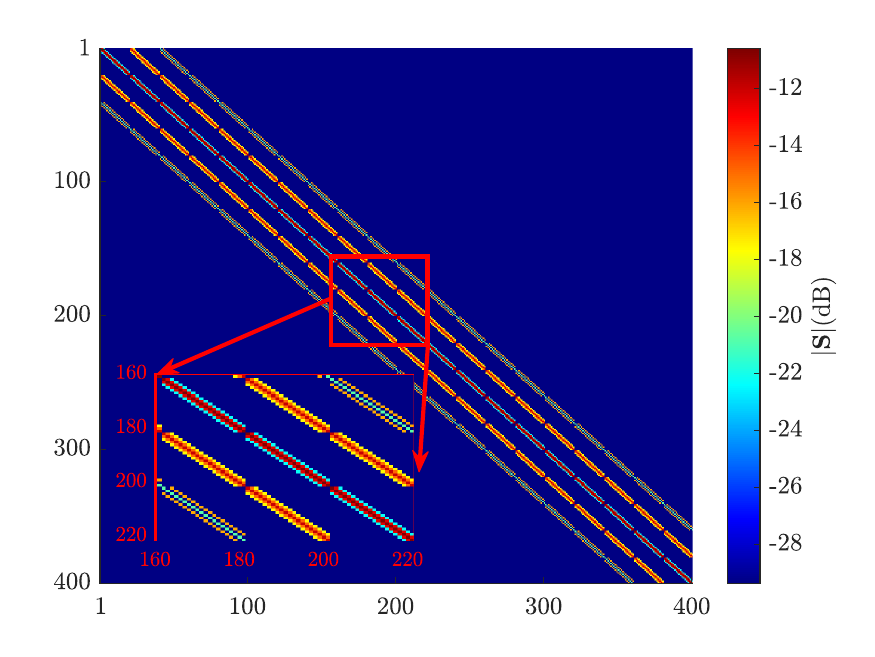}}
 \caption{Illustration of the scattering matrix sparse structure in the presence of \ac{mc} at the \ac{ris} side.}
 \label{fig:s_para}
\end{figure*}
\label{sec:CE_MC}

The primary distinction when formulating the exact and conventional sparse recovery problems is the structure of $\bar{\Gammam}$ (a full matrix) and $\Gammam$ (a diagonal matrix), where we omit the training index for notational simplicity. While the exact model provides highly accurate estimation, it is important to understand the propagation mechanism and incorporate the underlying physics of the~\ac{mc} since the severity of this effect is inversely proportional to element spacing. 

The~\ac{em} interactions among closely adjacent \ac{ris} unit cells are high; however, the more separated cells will experience much weaker interactions and, hence, can be ignored, as shown in Fig.~\ref{fig:mc_mech}. This observation hints that the scattering matrix $\Sm$ exhibits a sparse structure. To verify, we plot in Fig.~\ref{fig:real_s} the scattering matrix $\Sm$, estimated following the method developed in~\cite {Zheng2024Mutual}, for a real \ac{ris} prototype~\cite{Wang2024Wideband}. The analysis of~\cite{Zheng2024Mutual} considers only adjacent cell coupling and ignores weak coupling beyond two unit cell spacing. Note that the \ac{ris} prototype~\cite{Wang2024Wideband} deploys a cell spacing of $0.35\lambda_c$, therefore, the two-cell spacing was sufficient. For general scenarios, unit cells lying within a certain distance, such as one- or multiple-wavelength spacing, should be accounted for. However, the scattering matrix $\Sm$ is still expected to be sparse, since the \ac{mc} is primarily determined by the \ac{ris} unit cell physical layout.

To this end, since exploiting the sparsity of $\Sm$ in $\bar{\Gammam}$ is an intractable process due to the matrix inversion in~\eqref{eq:MC_ChM}, we apply the Neumann series expansion to approximate $\bar{\Gammam}$ and reformulate such inversion into tractable terms
\begin{equation}
\label{eq:Neumann}
    (\Gammam^{-1}\!\!-\Sm)^{-1}\!\!=\!\!\Big(\sum_{n=0}^\infty \big(\Gammam\Sm\big)^n\Big)\Gammam
    \!=\!\Gammam\! + \Gammam\Sm\Gammam + \Gammam\Sm\Gammam\Sm\Gammam\! + \cdots\!.
\end{equation}
The Neumann expansion is valid provided that all the eigenvalues of~$\Gammam\Sm$ are within the unit circle, i.e.,~$|\lambda_i(\Gammam\Sm)|<1$, $\forall\ i=1,\dots,N_\mathrm{I}$. Fortunately, such a condition can be satisfied, and the expression is valid for a typical~\ac{mc} strength. Therefore, we can truncate the Neumann series in~\eqref{eq:Neumann} to obtain a linear approximation of $(\Gammam^{-1} - \Sm)^{-1}$. A typical evaluation of the truncation error for different orders of expansion can be found in, e.g.,~\cite[Fig.~10]{Zheng2024Mutual}.
Retaining only the first two terms of~\eqref{eq:Neumann}, we write $(\Gammam^{-1}-\Sm)^{-1} \approx \Gammam + \Gammam\Sm\Gammam$ and define
\begin{align}
    \label{eq:neum_Gammabar}
    \bar{\Gammam}^\NEUM &=\Gammam + \Gammam\Sm\Gammam=\bar{\Gammam}^\APR \odot \Sm,\\
    \label{eq:approx_Gammabar}
    \bar{\Gammam}^\APR \!\!&=\!\!\begin{bmatrix} \frac{\gamma_1}{s_{1,1}}\!+\!\gamma_1^2&\gamma_1 \gamma_2 &\cdots&\gamma_1 \gamma_{N_{\Irm}} \\
    \gamma_1 \gamma_2 &\frac{\gamma_2}{s_{2,2}}\!+\!\gamma_2^2 &\cdots&\gamma_2 \gamma_{N_{\Irm}}\\
    \vdots&\vdots&\ddots&\vdots\\
    \gamma_1 \gamma_{N_{\Irm}} &\gamma_2 \gamma_{N_{\Irm}} &\cdots&\frac{\gamma_{N_{\Irm}}}{s_{N_{\Irm},N_{\Irm}}}\!+\!\gamma_{N_{\Irm}}^2
    \end{bmatrix}.
\end{align}
Therefore, $\bar{\Gammam}^\NEUM$ in~\eqref{eq:neum_Gammabar} has the same sparse structure as the matrix $\Sm$. Then, we can infer that the dictionary matrix $\tilde{\Am}_{\Irm\Irm}(\phiv,\thetav)$ will be multiplied by a sparse matrix; hence, we can select only the relevant rows of $\tilde{\Am}_{\Irm\Irm}^\Tran(\phiv,\thetav)$ from the dictionary matrix to perform the estimation. Moreover, we can leverage this sparsity to select the relevant elements of $\Thetam_\MC$ based on $\vec(\bar{\Gammam}^\NEUM)$. Another important note is that the~\ac{ris} amplitude and phase coefficients in~\eqref{eq:approx_Gammabar} change during training, so our design should be independent of the \ac{ris} configurations since we use random beams while collecting the measurements. 
\begin{algorithm}[htb!]
 \caption{\small{Proposed MC-aware sensing matrix design}}
 \label{algo:prop_sens_mat}
 \begin{algorithmic}[1]
 \Statex \textbf{Input:} $\Sm$, $\bar{a}$, $\epsilon_\THR$. \quad  \textbf{Output:} $\kappa_\PROP$,$\Psim_\PROP, \bar{\Dm}_\PROP$
 \State Construct the beam-scanning matrix $\Gammam_\SCN$ in~\eqref{eq:scanbeam_Gamma}
 \State Initialize the index set $\mathcal{I}= \{ \emptyset \}$
 \For {$n = 1,2,\cdots,N_\Irm$}
    \State Compute $\bar{\Gammam}^\NEUM$ of~\eqref{eq:neum_Gammabar} using $\Gammam_\TMP=\diag([\Gammam_\SCN]_{:,n})$
    \State Determine $\epsilon = \max_{_{1 \leq r,c \leq N_\Irm}}\Big(\abs{[\bar{\Gammam}^\NEUM]_{r,c}}\Big)/\epsilon_\THR$
    \State Find indices $(r, c)$ where $\abs{\bar{\Gammam}^\NEUM_{r,c}} \geq \epsilon$ and store in $\mathcal{I}_n$
 \EndFor
 \State Aggregate indices $\mathcal{I}_\PROP = \bigcup_{n=1}^{N_\Irm} \mathcal{I}_n$
 \State Calculate reduction ratio $\kappa_\PROP=\abs{\mathcal{I}_\PROP}/ N_\Irm^2 \times 100$
 \State Map linear indices $\mathcal{I}_\PROP$ to matrix-based row-column pairs $(\rv_\PROP, \cv_\PROP)$
 \State Initialize $\tilde{\Am}_{\Irm\Irm}^\PROP(\phiv,\thetav)\in\CC^{G_\Irm^2\times \abs{\mathcal{I}_\PROP}}$
 \For {$i = 1,2,\cdots,\abs{\mathcal{I}_\PROP}$}
     \State $[\tilde{\Am}_\PROP]_{:, i}(\phiv,\thetav) = [\tilde{\Am}_\Irm^\Tran(\phiv)]_{:,\cv_\PROP(i)}\otimes[\tilde{\Am}_\Irm^\Tran(\thetav)]_{:,\rv_\PROP(i)}$  
 \EndFor
 \State Compute $\bar{\Dm}_\PROP = \big(\tilde{\Am}_{\Irm\Irm}^\PROP\big)^\Tran(\phiv,\thetav)\otimes\tilde{\Am}_{\Trm\Rrm}(\varphiv,\varthetav)$ 
 \State Initialize $\Thetam_\PROP\in\CC^{\abs{\mathcal{I}_\PROP}\times M_\Irm}$
 \For {$m = 1,2,\cdots,M_\Irm$}
     \State $[\Thetam_\PROP]_{:,m} =[\Thetam_\MC]_{\mathcal{I}_\PROP,m}$
 \EndFor
 \State Compute $\Psim_\PROP=(\Thetam_\PROP^\Tran\otimes\Pm)$ 
 \end{algorithmic} 
\end{algorithm}

Following the previous analysis, we outline Algorithm~\ref{algo:prop_sens_mat} to obtain the proposed dictionary and measurement matrices that will be used for~\ac{cs}-based channel estimation of the received signal in~\eqref{eq:Rxsig_CS}. In step 1, we construct a beam-scanning matrix $\Gammam_\SCN\in\CC^{N_\Irm\times N_\Irm}$ where each column is
\begin{equation}
    \label{eq:scanbeam_Gamma}
    [\Gammam_\SCN]_{:,n}= \bar{a}e^{j\frac{(n-1)\pi}{N_\Irm}\nv(N_\Irm)}, n\in\{1,\cdots,N_\Irm\},
\end{equation}
and represents a beam covering an angular sector in the $[0,\pi)$ range, where each sector angular resolution is $\pi/N_\Irm$. After initializing the index set $\mathcal{I}$ as empty in step 2, we start an iterative loop to select the significant component indices that will impact the estimation (steps 3 to 6). For each column of~\eqref{eq:scanbeam_Gamma}, we compute $\Gammam_\TMP=\diag([\Gammam_\SCN]_{:,n})$ and utilize it to get $\bar{\Gammam}^\NEUM$ of~\eqref{eq:neum_Gammabar} (step 4). We define in step 5 the threshold level $\epsilon$ to be the ratio between the maximum element in $\bar{\Gammam}^\NEUM$ and a pre-determined threshold $\epsilon_\THR$, utilized to select the elements that are expected to influence the performance and filter out insignificant entries. This parameter will impact the size of the resultant dictionary and measurement matrices. After that, the indices that meet or exceed the threshold are stored in the set $\mathcal{I}_n$. In step 7, we aggregate the indices obtained from each iteration into the set $\mathcal{I}_\PROP$ that will be used to construct the sensing matrix. We compute the reduction ratio $\kappa_\PROP$ based on the cardinality of $\mathcal{I}_\PROP$, which represents the number of effective elements used while estimating the channel. Such a number is much less than $N_\Irm^2$ and is $\mathcal{O}(N_\Irm)$. After mapping back, in step 9, the linear indices of $\mathcal{I}_\PROP$ to row-column pairs $(\rv_\PROP, \cv_\PROP)$, we compute the proposed dictionary $\bar{\Dm}_\PROP$ in step 13 based on $\tilde{\Am}_\Irm(\phiv), \tilde{\Am}_\Irm(\thetav)$, and the selected row-column pairs from step 9. We utilize $(\rv_\PROP, \cv_\PROP)$ to select the relevant elements from $\Thetam_\PROP$ and then compute the measurement matrix $\Psim_\PROP$ (steps 14 to 17).\footnote{We assume that the \ac{bs} has prior knowledge of neighbor active \ac{ris} scattering matrix $\Sm$ and the average amplification factor $\bar{a}$, obtained through a side link. This assumption introduces minimal overhead, as it requires only knowledge of $\bar{a}$ and the index of the surrounding \ac{ris} with which the \ac{bs} is communicating. \cite{Zheng2024Mutual} proposes a model training approach to estimate $\Sm$ using a single 3D full-wave simulation of the \ac{ris} radiation pattern, which can be conducted once and stored for future use. $\Sm$ is predominantly determined by the \ac{ris} physical architecture and is therefore quasi-static over the operating period; however, calibration/modeling errors and hardware-induced drift can cause deviations from the nominal stored matrix. The effect of scattering-matrix mismatches is left as future work.} To this end, by using the \ac{omp} on-grid CS algorithm \cite{Lee2016Channel}, we can estimate the sparse vector $\bar{\sigmav}$ in \eqref{eq:Rxsig_CS} depending on the received signal and a suitable sensing matrix $\Xim_\chi\!=\!\Psim_\chi\bar{\Dm}_\chi\!\in\!\CC^{M_\Rrm M_\Irm \times \bar{G}_\chi}$ where~$\chi\in\{\CV,\PROP,\MC\}$. It is expected that the conventional model-based solution suffers from significant model mismatch due to neglecting the \ac{mc}. In contrast, the exact model-based solution is accurate but involves a dimensional lift. However, our proposed sensing matrix design enjoys the advantages of both methods with reduced complexity compared to the exact model approach.

\section{Simulation results and discussion}
\label{sec:Sim_Res_Disc}
\begin{figure}[t]
    \centering
    \begin{minipage}[b]{0.98\linewidth}
    \centering
%
%
\definecolor{mycolor1}{rgb}{1.00000,0.00000,1.00000}%
\begin{tikzpicture}

\begin{axis}[%
width=2.6in,
height=1.7in,
at={(0in,0in)},
scale only axis,
xmin=-7,
xmax=15,
xlabel style={font=\color{white!15!black},font=\footnotesize},
xtick={-5,0,5,10,15},
xticklabel style = {font=\color{white!15!black},font=\footnotesize},
xlabel={Transmit power $P_\Trm$ (dBm)},
ymin=-30,
ymax=12,
ylabel style={font=\color{white!15!black},font=\footnotesize},
ytick={-30,-20,-10,0,10},
yticklabel style = {font=\color{white!15!black},font=\footnotesize},
ylabel={Estimation NMSE (dB)},
axis background/.style={fill=white},
xmajorgrids,
ymajorgrids,
grid style={dotted, black, opacity=1},
legend style={
    at={(0,0)}, 
    anchor=south west, 
    legend cell align=left, 
    align=left, 
    draw=white!15!black, 
    font=\scriptsize,
    inner sep=2pt,
    row sep=-1pt,
    nodes={inner sep=1pt},
    fill opacity=0.85
}
]
\addplot [color=green, line width=1pt, mark size=3pt, mark=triangle, mark options={solid, rotate=90, green}]
  table[row sep=crcr]{%
-6	10.2678220748901\\
-4	7.60317425727844\\
-2	4.95676426887512\\
0	2.87757532596588\\
2	1.39977064728737\\
4	0.194087316840887\\
6	-0.695522075891495\\
8	-1.23784438371658\\
10	-1.63143136501312\\
12	-1.84832229614258\\
14	-2.06291670799255\\
};
\addlegendentry{MC-unaware}

\addplot [color=blue, line width=1pt, mark size=3pt, mark=triangle, mark options={solid, rotate=270, blue}]
  table[row sep=crcr]{%
-6	8.24606947898865\\
-4	4.64531158208847\\
-2	1.20111760497093\\
0	-1.82769193649292\\
2	-4.6384138584137\\
4	-6.74741930961609\\
6	-8.52307324409485\\
8	-10.0477844238281\\
10	-10.9491676330566\\
12	-11.7235074996948\\
14	-12.0638974189758\\
};
\addlegendentry{Prop. $\kappa_{\mathrm{prop}} = 6.18\%$}

\addplot [color=red, line width=1pt, mark size=3.3pt, mark=diamond, mark options={solid, red}]
  table[row sep=crcr]{%
-6	7.84017729759216\\
-4	4.21783339977264\\
-2	0.254550978913903\\
0	-2.93539090156555\\
2	-6.4239236831665\\
4	-9.17041778564453\\
6	-12.0322657585144\\
8	-14.1908005714417\\
10	-15.9905102729797\\
12	-17.2245714187622\\
14	-17.8400249481201\\
};
\addlegendentry{Prop. $\kappa_{\mathrm{prop}} = 13.31\%$}

\addplot [color=mycolor1, line width=1pt, mark size=3pt, mark=asterisk, mark options={solid, mycolor1}]
  table[row sep=crcr]{%
-6	7.78441395759583\\
-4	3.79187990427017\\
-2	-0.274773164093494\\
0	-4.0008266210556\\
2	-7.91976914405823\\
4	-11.6015340805054\\
6	-14.9653221130371\\
8	-17.9399936676025\\
10	-20.8349813461304\\
12	-22.840541267395\\
14	-24.6632120132446\\
};
\addlegendentry{Prop. $\kappa_{\mathrm{prop}} = 25.27\%$}
\addplot [color=black, line width=1pt, mark size=2.3pt, mark=square, mark options={solid, black}]
  table[row sep=crcr]{%
-6	7.85514831542969\\
-4	3.77249867022037\\
-2	-0.248617321252823\\
0	-4.22330932617187\\
2	-8.33663792610168\\
4	-11.6592284202576\\
6	-15.6264102935791\\
8	-19.1000602722168\\
10	-22.0839370727539\\
12	-25.2949989318848\\
14	-27.5290994644165\\
};
\addlegendentry{MC-aware}

\end{axis}

\end{tikzpicture}%
    \vspace{-1.cm}
    \centerline{\footnotesize{(a) $N_\Irm=16\times8$, $N_\Rrm=2\times2$, and $N_\Trm=2\times2$}} \medskip
    \end{minipage}
    \hfill
    \begin{minipage}[b]{0.98\linewidth}
    \centering
%
%
\definecolor{mycolor1}{rgb}{1.00000,0.00000,1.00000}%
\begin{tikzpicture}

\begin{axis}[%
width=2.6in,
height=1.7in,
at={(0in,0in)},
scale only axis,
xmin=-16.3,
xmax=0.3,
xlabel style={font=\color{white!15!black},font=\footnotesize},
xticklabel style = {font=\color{white!15!black},font=\footnotesize},
xlabel={Transmit power $P_\Trm$ (dBm)},
ymin=-21.8039264678955,
ymax=13.9273118972778,
ylabel style={font=\color{white!15!black},font=\footnotesize},
yticklabel style = {font=\color{white!15!black},font=\footnotesize},
ylabel={Estimation NMSE (dB)},
axis background/.style={fill=white},
xmajorgrids,
ymajorgrids,
grid style={dotted, black, opacity=1},
legend style={
    at={(0,0)}, 
    anchor=south west, 
    legend cell align=left, 
    align=left, 
    draw=white!15!black, 
    font=\scriptsize,
    inner sep=2pt,
    row sep=-1pt,
    nodes={inner sep=1pt},
    fill opacity=0.85
}
]
\addplot [color=green, line width=1pt, mark size=3pt, mark=triangle, mark options={solid, rotate=90, green}]
  table[row sep=crcr]{%
-16	12.3031648635864\\
-14	9.39539318084717\\
-12	6.79734826087952\\
-10	4.61215181350708\\
-8	2.84566164016724\\
-6	1.44700254201889\\
-4	0.614851686358452\\
-2	-0.0625876891426742\\
0	-0.506890866160393\\
};
\addlegendentry{MC-unaware}

\addplot [color=blue, line width=1pt, mark size=3pt, mark=triangle, mark options={solid, rotate=270, blue}]
  table[row sep=crcr]{%
-16	10.8361682891846\\
-14	7.25874800682068\\
-12	3.90690703392029\\
-10	0.891090319305658\\
-8	-1.74041336774826\\
-6	-3.87676155567169\\
-4	-5.49528317451477\\
-2	-6.71269326210022\\
0	-7.62505354881287\\
};
\addlegendentry{Prop. $\kappa_{\mathrm{prop}} = 3.23\%$}

\addplot [color=red, line width=1pt, mark size=3.3pt, mark=diamond, mark options={solid, red}]
  table[row sep=crcr]{%
-16	10.539884185791\\
-14	6.66443948745728\\
-12	2.92720417976379\\
-10	-0.789913830161095\\
-8	-4.10760383605957\\
-6	-7.16687426567078\\
-4	-9.94229011535645\\
-2	-12.0716344833374\\
0	-13.7864909172058\\
};
\addlegendentry{Prop. $\kappa_{\mathrm{prop}} = 7.16\%$}

\addplot [color=mycolor1, line width=1pt, mark size=3pt, mark=asterisk, mark options={solid, mycolor1}]
  table[row sep=crcr]{%
-16	10.8863943099976\\
-14	6.93383779525757\\
-12	3.20416009426117\\
-10	-0.55496208127588\\
-8	-4.71070461273193\\
-6	-8.45943074226379\\
-4	-11.6440614700317\\
-2	-14.4385722160339\\
0	-17.0555559158325\\
};
\addlegendentry{Prop. $\kappa_{\mathrm{prop}} = 14.2\%$}
\addplot [color=black, line width=1pt, mark size=2.3pt, mark=square, mark options={solid, black}]
  table[row sep=crcr]{%
-16	10.998595905304\\
-14	6.86825604438782\\
-12	3.07093753814697\\
-10	-0.814787121117115\\
-8	-5.1907632112503\\
-6	-9.01086440086365\\
-4	-12.8656644821167\\
-2	-16.5858761787415\\
0	-20.179779624939\\
};
\addlegendentry{MC-aware}

\end{axis}

\end{tikzpicture}%
    \vspace{-1.cm}
    \centerline{\footnotesize{(b) $N_\Irm=16\times16$, $N_\Rrm=2\times2$, and $N_\Trm=1\times1$}} \medskip
    \end{minipage}
    \caption{Performance evaluation of estimation accuracy versus transmit power assessed by NMSE.}
    \label{fig:chest_perf}
    \vspace{-1.5em}
\end{figure}
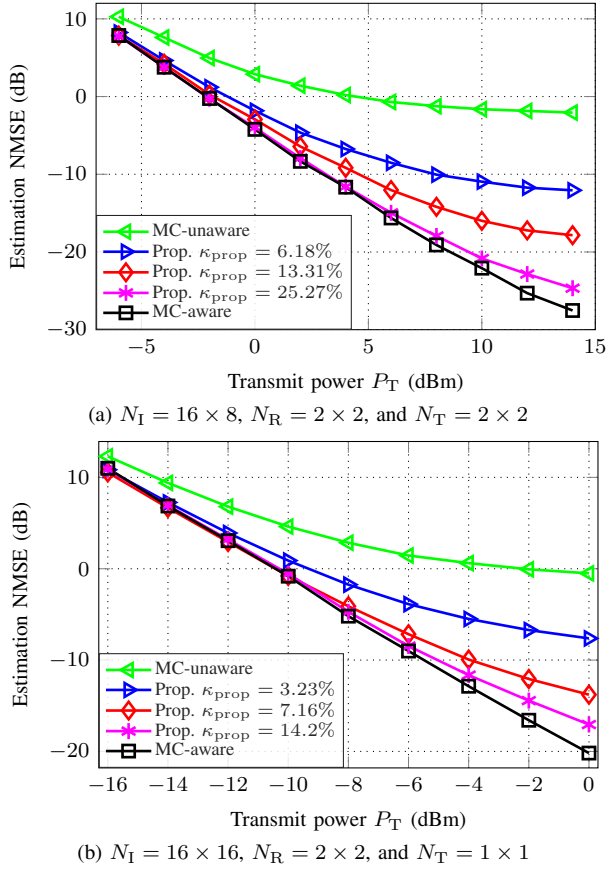

For a fair comparison, since the estimated $\hat{\Gm}_\CV$ and~$\hat{\Gm}_\MC$ have different dimensions, we evaluate the estimation accuracy by computing the \ac{nmse} metric of the reconstructed received signal as  
\begin{equation}
    \text{NMSE}_\chi = \EE\|\Pm\hat{\Gm}_\chi\Thetam_\chi-\bar{\Ym}\|_\mathrm{F}^2/\|\bar{\Ym}\|_\mathrm{F}^2,
\end{equation}
where~$\bar{\Ym}$ is the noise-free version of the exact-model received signal. This is motivated by the fact that the estimation channels have different dimensions. We define $P_\Trm$ as the total \ac{ue} power used per transmission during the training phase.

Throughout the simulation, the \ac{ris} is at~$[0, 0, 0]^\Tran$ facing the positive Y-axis, the \ac{ue} is at $\unit[2.6]{m} \times [\sin{\frac{\pi}{6}}, \cos{\frac{\pi}{6}}, 0]^\Tran$ facing the negative Y-axis, and the \ac{bs} at~$\unit[2.2]{m} \times [\sin{\frac{\pi}{3}}, \cos{\frac{\pi}{3}}, 0]^\Tran$ facing the negative Y-axis, consistent with the measurement setup in~\cite{Wang2024Wideband}. The~\ac{ris} has $N_\Irm=16\times8$ unit cells with $d_\Irm=\lambda_c/20$, while the~\ac{bs} and~\ac{ue} deploy $N_\Rrm=2\times2$, $N_\Trm=2\times2$ elements, respectively, with $d_\Rrm=d_\Trm=\lambda_c/2$. Unless stated otherwise, we set the~\ac{ris} average amplification factor~$\bar{a}\!=\!6$. The training uses uniformly distributed random phase shifts $\mathcal{U}[0,2\pi)$. We perform $100$ trials to average the results and set $G_\Irm\!=\!N_\Irm$, $G_\Rrm\!=\!N_\Rrm$, $G_\Trm\!=\!N_\Trm$, $M_\Irm\!=\!N_\Irm/2$, and $M_\Rrm\!=\!N_\Rrm$. Note that during the simulation, the number of~\ac{omp} iterations is $\hat{L}$, which is the number of paths to be estimated and is assumed to be different from the actual channel sparsity level, since it is an unknown parameter. We set $f_c=\unit[30]{GHz}$ and the thermal noise power~$\sigma_\Rrm^2=\sigma_\Irm^2=\unit[-95]{dBm}$. Regarding channel generation, we use $L_{\Trm}=4$, the \ac{los} path gain in~\eqref{eq:ch_TxRIS_spatial_domain} is $\alpha_1=(\lambda_c/(4\pi d_{\Irm\Trm}))^{\gamma_{\Irm\Trm}/2}$, where $d_{\Irm\Trm}$ is the distance between~\ac{ue} and~\ac{ris}, and the path loss exponent $\gamma_{\Irm\Trm}=2.1$. For \ac{nlos} paths $\alpha_\ell\sim\mathcal{CN}(0,\alpha_1^2)$. The same logic applies to $\rho_\ell$ in~\eqref{eq:ch_RISRx_spatial_domain} with $L_\Rrm=2$. The scattering parameters are calculated similarly to \cite{Zheng2026Mutual}.

Fig.~\ref{fig:chest_perf} presents the estimation \ac{nmse} versus transmit power. We set the linear parameter $\epsilon_\THR$ so that its values in dB are $\in\{10, 15,20\}$ in Algorithm~\ref{algo:prop_sens_mat}, leading to dictionary and measurement matrices of size $\kappa_\PROP \in\{6.18\%,13.31\%,25.27\%\}$ (Fig.~\ref{fig:chest_perf}(a)) and $\kappa_\PROP \in\{3.29\%,7.16\%,14.2\%\}$ (Fig.~\ref{fig:chest_perf}(b)) of that required by the exact-model approach, respectively. As demonstrated, the proposed algorithm consistently outperforms the conventional MC-unaware OMP in the presence of \ac{mc} achieving several dB improvements in estimation accuracy. Additionally, the estimation accuracy increases by accounting for more significant elements. The proposed method slightly differs from the MC-aware exact model. However, a huge reduction in complexity is obtained while still achieving comparable performance.

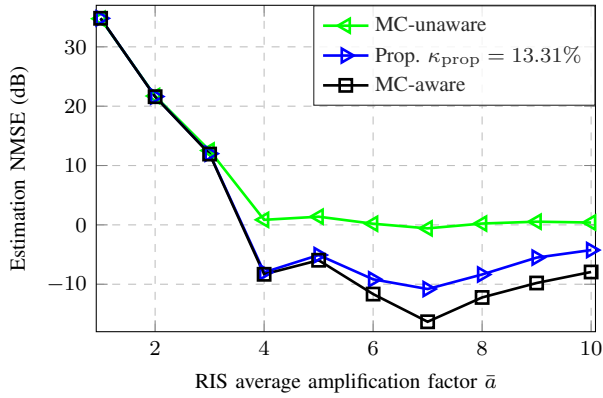
\begin{figure}[htb]
    \centering
%
\definecolor{ForestGreen}{rgb}{0.1333    0.8451    0.1333}
\begin{tikzpicture}

\begin{axis}[%
width=2.6in,
height=1.7in,
at={(0in,0in)},
scale only axis,
xmin=0.92,
xmax=10.08,
xminorticks=true,
xlabel style={font=\color{white!15!black},font=\footnotesize},
xticklabel style = {font=\color{white!15!black},font=\footnotesize},
xlabel={RIS average amplification factor~$\bar{a}$},
ymin=-18,
ymax=37,
ylabel style={font=\color{white!15!black},font=\footnotesize},
yticklabel style = {font=\color{white!15!black},font=\footnotesize},
ytick = {-20,-10,0,10,20,30},
ylabel={Estimation NMSE (dB)},
axis background/.style={fill=white},
xmajorgrids,
xminorgrids,
ymajorgrids,
grid style={dashed},
legend style={at={(1,1)}, anchor=north east, font=\footnotesize, legend cell align=left, align=left, draw=white!15!black, fill opacity=0.85}
]
\addplot [color=green, line width=1pt, mark size=3pt, mark=triangle, mark options={solid, rotate=90, green}]
  table[row sep=crcr]{%
1	34.7400508880615\\
2	21.7135454177856\\
3	12.5327684402466\\
4	0.851629756391048\\
5	1.37985846996307\\
6	0.194087316840887\\
7	-0.589402908086777\\
8	0.226158792339265\\
9	0.536433044075966\\
10	0.393800781667233\\
};
\addlegendentry{MC-unaware}

\addplot [color=blue, line width=1pt, mark size=3pt, mark=triangle, mark options={solid, rotate=270, blue}]
  table[row sep=crcr]{%
1	34.805424118042\\
2	21.6313926696777\\
3	12.0004319190979\\
4	-7.99992170333862\\
5	-5.08434827327728\\
6	-9.17041778564453\\
7	-10.8107267379761\\
8	-8.35843629837036\\
9	-5.51699166297913\\
10	-4.24497125148773\\
};
\addlegendentry{Prop. $\kappa_{\mathrm{prop}} = 13.31\%$}

\addplot [color=black, line width=1pt, mark size=2.3pt, mark=square, mark options={solid, black}]
  table[row sep=crcr]{%
1	34.813667678833\\
2	21.6001056671143\\
3	11.9320178985596\\
4	-8.32588686943054\\
5	-5.95901689529419\\
6	-11.6592284202576\\
7	-16.3261076927185\\
8	-12.2286805152893\\
9	-9.79490051269531\\
10	-7.93732261657715\\
};
\addlegendentry{MC-aware}

\end{axis}

\end{tikzpicture}%
    \vspace{-3em}
    \caption{Performance evaluation of the estimation accuracy versus~\ac{ris} average amplification factor~$\bar{a}$.}
    \label{fig:nmse_vs_rispower}
    \vspace{-1em}
\end{figure}
\begin{figure}[htb]
    \centering
%
\definecolor{ForestGreen}{rgb}{0.1333    0.8451    0.1333}
\begin{tikzpicture}

\begin{axis}[%
width=2.6in,
height=1.7in,
at={(0in,0in)},
scale only axis,
xmode=log,
xmin=0.02,
xmax=0.5,
xminorticks=true,
xlabel style={font=\color{white!15!black},font=\footnotesize},
xticklabel style = {font=\color{white!15!black},font=\footnotesize},
xtick = {1/50,1/32,1/16,1/8,1/4,1/2},
xticklabels = {$\frac{\lambda}{50}$,$\frac{\lambda}{32}$,$\frac{\lambda}{16}$,$\frac{\lambda}{8}$,$\frac{\lambda}{4}$,$\frac{\lambda}{2}$},
xlabel={RIS inter-element spacing $d_\Irm$},
ymin=-17,
ymax=10,
ylabel style={font=\color{white!15!black},font=\footnotesize},
yticklabel style = {font=\color{white!15!black},font=\footnotesize},
ytick = {10,5,0,-5,-10,-15},
ylabel={Estimation NMSE (dB)},
axis background/.style={fill=white},
xmajorgrids,
xminorgrids,
ymajorgrids,
grid style={dashed},
legend style={at={(1,0)}, anchor=south east, font=\footnotesize, legend cell align=left, align=left, draw=white!15!black, fill opacity=0.85}
]
\addplot [color=green, line width=1pt, mark size=3pt, mark=triangle, mark options={solid, rotate=90, green}]
  table[row sep=crcr]{%
0.02	1.17427411079407\\
0.03125	1.59264367818832\\
0.0364540324867536	1.54803954362869\\
0.0425246875054493	2.06818565130234\\
0.0496062828740062	0.0175114210695028\\
0.0578671695179556	-0.942337369918823\\
0.0675037336807691	3.9825963973999\\
0.0787450656184295	5.00887310504913\\
0.0918584057672249	5.75959296226501\\
0.107155497856634	5.85958151817322\\
0.125	7.24847464561462\\
0.125	7.24847464561462\\
0.198425131496025	6.80488309860229\\
0.314980262473718	4.03193454742432\\
0.5	8.7504301071167\\
};
\addlegendentry{MC-unaware}

\addplot [color=blue, line width=1pt, mark size=3pt, mark=triangle, mark options={solid, rotate=270, blue}]
  table[row sep=crcr]{%
0.02	-9.40799474716187\\
0.03125	-6.28025898933411\\
0.0364540324867536	-4.7421781539917\\
0.0425246875054493	-2.65203837156296\\
0.0496062828740062	-9.29716072082519\\
0.0578671695179556	-13.742712020874\\
0.0675037336807691	2.64708016514778\\
0.0787450656184295	4.72505521774292\\
0.0918584057672249	5.71288952827454\\
0.107155497856634	6.01139984130859\\
0.125	7.16973352432251\\
0.125	7.16973352432251\\
0.198425131496025	6.80462584495544\\
0.314980262473718	4.03186640739441\\
0.5	8.75041790008545\\
};
\addlegendentry{Prop. $\kappa_{\mathrm{prop}} = 13.31\%$}

\addplot [color=black, line width=1pt, mark size=2.3pt, mark=square, mark options={solid, black}]
  table[row sep=crcr]{%
0.02	-9.65389590263367\\
0.03125	-9.51333575248718\\
0.0364540324867536	-8.62523684501648\\
0.0425246875054493	-7.1239296913147\\
0.0496062828740062	-11.2984630584717\\
0.0578671695179556	-15.1379029273987\\
0.0675037336807691	2.37481752336025\\
0.0787450656184295	4.97988474369049\\
0.0918584057672249	5.68077545166016\\
0.107155497856634	5.9662914276123\\
0.125	7.32547216415405\\
0.125	7.32547216415405\\
0.198425131496025	6.83989396095276\\
0.314980262473718	4.01176991462708\\
0.5	8.76980214118958\\
};
\addlegendentry{MC-aware}

\end{axis}

\end{tikzpicture}%
    \vspace{-3em}
    \caption{Performance evaluation of the estimation accuracy versus~\ac{ris} inter-element spacing.}
    \label{fig:nmse_vs_spacing}
    \vspace{-1em}
\end{figure}
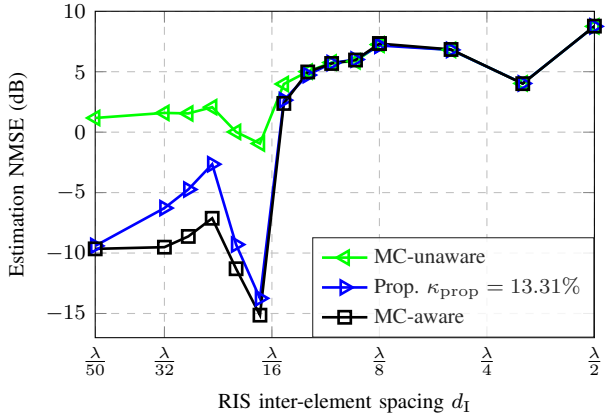

Fig.~\ref{fig:nmse_vs_rispower} evaluates the impact of the active \ac{ris} power constraint, by adjusting the amplitude weights of the \ac{ris} elements, on the estimation accuracy. The results demonstrate that higher \ac{ris} amplification intensifies \ac{mc} effects, causing severe performance degradation in conventional \ac{mc}-unaware models. In contrast, the proposed method exhibits strong robustness to varying \ac{mc} levels, consistently matching the accuracy of the exact \ac{mc}-aware formulation while requiring significantly lower computational complexity.
In Fig. \ref{fig:nmse_vs_spacing}, we examine different \ac{mc} levels by adjusting the RIS inter-element spacing. At low element separation (corresponding to higher \ac{mc} levels), the conventional model shows significant degradation, while at larger element spacings, i.e., negligible \ac{mc} effects, its performance approaches that of the exact model. The proposed method consistently provides performance similar to the exact model while changing~\ac{ris} element spacing.

\section{Conclusion}
\label{sec:conclusion}
This paper addresses the challenges of channel estimation in active \ac{ris}-assisted \ac{mimo} communication systems in the presence of \ac{mc}. By exploiting the \ac{mc} mechanism and the Neumann series expansion, combined with the \ac{cs} techniques, we propose a low-complexity yet accurate channel estimator that performs effectively under strong \ac{mc} conditions. Specifically, we exploit the fact that even for densely packed \ac{ris}, the unit cells lying very far away from a reference unit cell experience a reduced \ac{mc} effect, resulting in a sparse structure of the sensing matrix used for \ac{cs} problem formulation.

\bibliography{abbrev,references}
\bibliographystyle{IEEEtran}

\end{document}